# A thought to illustrate the uncertainty principle on the base of Otto-Wiener's experiment

**Tapas Das**
Kodalia Prasanna Banga High School (H.S), South 24 Parganas, 700146, India
E-mail: tapasd20@gmail.com

**Abstract**
In this short paper a new thought experiment has been introduced to illustrate the famous Heisenberg's uncertainty principle based on Otto-Wiener's experiment (1890) associated with standing light waves. This illustration is quite easy as well as far more realizing than all other thought experiments generally found in textbooks. In this work seeding of quantum nature of light has been done with the Otto-Wiener's experiment. May be this new thought experiment will help the students to understand the Heisenberg's principle in a better way and also enhance their interest of learning quantum mechanics from the beginning of their course.

**1. Introduction**
Heisenberg's uncertainty principle [1] is one of the important principles in quantum mechanics. We use this fascinating principle not only for non-relativistic quantum mechanics, but also for particle physics, nuclear physics as well as for hadronic study. Over the decades physicists in this field tried to illustrate Heisenberg's uncertainty principle with various thought experiments. The first high precision experimental test for the uncertainty relations came out only in 1969 from Shull's single slit neutron diffraction experiment [2]. Later in the 1980s followed the neutron interferometric experiments by Kaiser *et al* [3] and Klien *et al* [4]. Few years back in 2002 Nairz *et al* [5] have reported a demonstration of uncertainty relation by measuring the increase in momentum spread of the molecular beam of fullerene ($C_{70}$) molecules after passage through a narrow slit with a variable width. An excellent update status of uncertainty relation, can be found in the reference [6].
Generally most textbooks [7-9] of quantum mechanics introduce gamma ray microscope based thought experiment or particle diffraction based thought experiment for beginner's learning. Recently, Shivalingaswamy and Kagali [10] introduced a new thought experiment based on gamma ray telescope which is far more better than the thought experiment based on gamma ray microscope. It is worth to mention here that above said thoughts are based on the gamma ray, which are not so beginner friendly with respect to realization of the fact. It is time to switch from "*Gamma ray based thoughts*" to a new thought based on simple optical experiment which is easy to demonstrate for classroom physics. In this paper Otto-Wiener's historical experiment [11] on standing light waves has been introduced for easy illustration of Heisenberg's uncertainty principle. Basically this optical experiment is a good example of the superposition of electromagnetic field.

## 2. A brief history and summary of the Wiener's experiment

In 1865, James Clerk Maxwell theoretically demonstrated electromagnetic wave equation based on electromagnetic field equations. The solution of the wave equation in free space claimed that light is an electromagnetic wave. Heinrich Hertz was first person who experimentally demonstrated in 1987 that electromagnetic waves were consistent with Maxwell's theory, measuring their velocity, electric field intensity and polarization properties. He also produced standing radio waves by reflection from a zinc plate. Before going into the further historical background it is important to mention about the physical meaning of standing wave briefly.

Standing waves are produced whenever two waves of an identical frequency traveling in opposite directions superpose with each other. Standing waves production for sound waves [$wave\ length\ \lambda \approx$ few meter] is very easy in common laboratory, but for light waves it is quite difficult to produce because of its extremely short wave length [ $wave\ length\ \lambda \approx 5 \times 10^{-5}\ cm$ ]. Standing wave patterns are characterized by certain fixed points (or planes) where the fields undergo no displacement with time. These points are called nodes. Midway between every consecutive nodal point there exists a point which undergoes maximum displacement. These points are called antinodes. Antinodes are the points along the medium that oscillate most. Unlike traveling waves, standing waves transmit no energy. All the energy in the waves goes into sustaining the oscillations between the nodes, at which forward and reverse wave exactly cancel each other.

Now back to the history again. Soon after Hertz's experiment the idea of electromagnetic nature of light opened a wide range of research aspect in literature specially in optics. In 1890 Otto-Wiener [11] first performed the experiment of standing wave of light. His arrangement is illustrated in figure-1. A plane mirror M, silvered on the front surface, was illuminated normally by a parallel beam of quasi-monochromatic light coming from a sodium arc lamp. A film F of transparent photographic emulsion, coated on the plane surface of a glass plate G and less than $\frac{1}{20}$ wavelength thick, was placed in front of M and inclined to it at a small angle $\alpha$. After the emulsion had been developed, he found that it was blackened in equidistant parallel bands, which was due to the formation of nodes and antinodes of the standing light wave. With normal incidence and a convex spherical reflecting surface pressed in contact with the emulsion, he was able to conclude that the fringes were due to the electric field vector alone. His conclusion supports the well known fact that the photochemical action is directly related to the electric and not to the magnetic vector. An excellent account of his experiment was introduced in textbooks on optics [12,13].

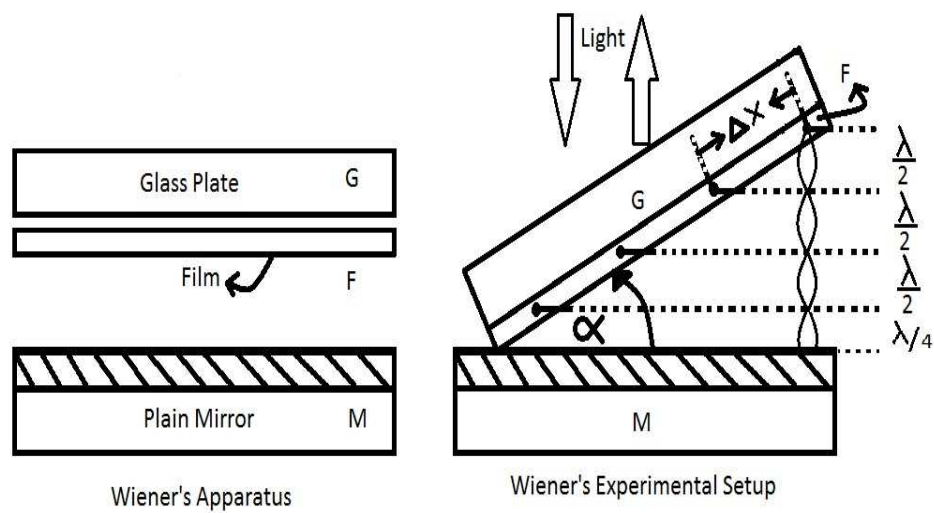

**Figure 1:** Left figure shows the apparatus of Otto-Wiener's experiment. Right figure shows the actual Otto-Wiener experimental setup diagram, where emulsion side of the photographic plate is placed with an small angle $\alpha$.

## 3. Uncertainty in position and momentum

Let us study the Otto-Wiener experiment via photon concept. Suppose a parallel beam of quasi-monochromatic light of wave length $\lambda$ is allowed to fall on the Otto-Wiener arrangement. Now according to quantum mechanics light can be thought of collection of photons. Standing wave between the incident and reflected light (or photon) will create dark bands in the transparent photographic emulsion. The separation between two dark bands can be taken as the uncertainty of position of the photon along the reference X-direction. So here we have

$$\Delta x = \frac{\lambda}{2 \sin\alpha}. \qquad (1)$$

Now it is time to think about the momentum components of the incident and reflected photons along the reference X-axis. The momentum $p = \frac{h}{\lambda}$ of the incident and reflected photons are same in magnitude, since the frequency of the light remains same due to reflection effect. Here $h$ denotes the famous Planck's constant. The direction of momentum are exactly opposite to each other. The momentum component along negative X-axis of the incident photon is $p_- = \frac{h}{\lambda}\sin\alpha$ whereas the same for the reflected photon along the positive X-axis is $p_+ = \frac{h}{\lambda}\sin\alpha$.

So we can think that, to sustain the vibration of standing wave these two momentum components are responsible.

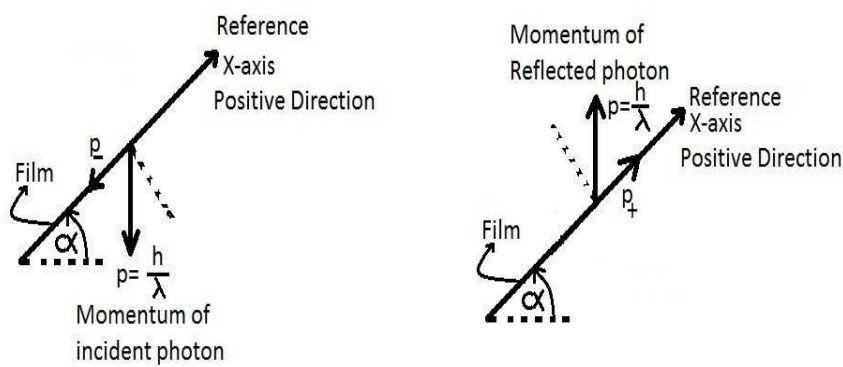

**Figure 2:** Calculation of uncertainty in momentum for incident (Left figure) and reflected (right figure) light.

Hence the uncertainty in momentum can be taken along reference X-axis as

$$\Delta p_x = p_- + p_+ = 2\frac{h}{\lambda}\sin\alpha. \qquad (2)$$

So multiplying Eq.(1) and Eq.(2) we have

$$\Delta x \Delta p_x = \frac{\lambda}{2\sin\alpha} 2\frac{h}{\lambda}\sin\alpha = h, \qquad (3)$$

which is the famous Heisenberg's uncertainty principle.

## 4. Discussion

This paper illustrates the Heisenberg's Uncertainty principle with the help of Otto-Wiener's historical experiment on standing wave of light. The result that has been presented here is independent of the wavelength of the light used as well as the parameter associated with the experimental apparatus, as it should. This new thought experiment will surely help the students as well as other readers, who want to begin first course in quantum mechanics. I would like to call this new thought as *Standing wave of light based thought experiment.*